\documentclass{phb-proc4-auth}

\usepackage{graphicx}
\usepackage{amssymb}
\usepackage{epsfig}
\usepackage{bm}
\usepackage{multicol}

\begin{document}
\begin{frontmatter}

\journal{Physics Letters A}

\title{Artificial scaling laws of the dynamical magnetic susceptibility in heavy-fermion systems}

\author{W. Knafo, S. Raymond}

\address{CEA-Grenoble, DSM/DRFMC/SPSMS/MDN, 38054 Grenoble
Cedex 9, France}

\begin{abstract}

We report here how artificial, thus erroneous, scaling laws of the
dynamical magnetic susceptibility can be obtained when data are
not treated carefully. We consider the example of the
heavy-fermion system Ce$_{0.925}$La$_{0.075}$Ru$_{2}$Si$_{2}$ and
we explain how different kinds of artificial scaling laws in
$E/T^\beta$ can be plotted in a low temperature regime where the
dynamical susceptibility is nearly temperature independent.

\end{abstract}


\end{frontmatter}

\section{Introduction}

In heavy-fermion systems, a magnetic instability can be reached by
tuning a parameter $\delta$ such as pressure, magnetic field or
doping. This instability is obtained at a critical value
$\delta_c$ of the tuning parameter and it separates a paramagnetic
ground state for $\delta<\delta_c$ from a magnetically ordered
ground state for $\delta>\delta_c$: a quantum phase transition
(defined at $T=0$) between those two states is thus induced at
$\delta_c$. While far from $\delta_c$ the paramagnetic state is
well described by Landau's Fermi liquid theory with a
strongly-renormalized effective mass $m^*\gg m_0$ ($m_0$ is the
free electron mass), a non-Fermi liquid behavior is generally
reported at the magnetic instability. In this regime, the specific
heat, the susceptibility and the resistivity follow power or
logarithmic temperature laws, which are different from the laws
expected for a Fermi liquid \cite{Stewartrevue}. This non-Fermi
liquid behavior is probably related both to the
$\mathbf{Q}$-dependence of the excitation spectra and to the
effects of temperature \cite{Hertz,Millis,Moriya}. However, this
regime is not yet explained for heavy fermions, its understanding
being now one of the major stakes in the physics of strongly
correlated electron systems. Recently, several inelastic neutron
scattering studies of heavy-fermion systems at their quantum phase
transition were performed with the aim to obtain scaling laws of
the dynamical magnetic susceptibility $\chi(E,T)$, where $E$ is
the energy transfer and $T$ the temperature
\cite{Aronson95,Schroder00,Park02a,Park02b,Montfrooij03,So03}. In
those works, the imaginary part $\chi''(E,T)$ of the dynamical
susceptibility was found to follow scaling laws of the form:
\begin{eqnarray}
  T^{\alpha}\chi''(E,T)=f(E/T)
  \label{scal}
\end{eqnarray}
down to $T=0$, with $\alpha<1$. These laws were interpreted as a
consequence of the divergence of the magnetic excitations when
$T\rightarrow0$ and $\delta\rightarrow\delta_c$, which is
responsible for the non-Fermi liquid behavior of these critical
systems. For example, the study carried out by Schr\"{o}der et al.
on CeCu$_{5.9}$Au$_{0.1}$ led to the conclusions that a general
scaling law of the form (\ref{scal}) is obtained down to $T=0$
with a unique exponent $\alpha=0.75$ for each wavevector of the
reciprocal space \cite{Schroder00}: this work motivated several
new theoretical developments, such as models based on a local
criticality \cite{Si,Coleman}.

In this letter, we show how it is possible to obtain artificial
scaling laws of the dynamical magnetic susceptibility when the
data are not carefully analyzed. We consider already published
data corresponding to the antiferromagnetic fluctuations of the
heavy-fermion system Ce$_{1-x}$La$_{x}$Ru$_{2}$Si$_{2}$ at its
magnetic instability $x_{c}=7.5\%$ \cite{Knafo1,Knafo2,Knafo3}. We
explain how erroneous scaling laws can be obtained in a low
temperature regime where the dynamical magnetic susceptibility
depends weakly on $T$. In such a nearly T-independent regime,
$E/T$ scaling laws are irrelevant.

\section{Antiferromagnetic fluctuations in Ce$_{0.925}$La$_{0.075}$Ru$_{2}$Si$_{2}$}

\begin{figure}[b!]
    \centering
    \epsfig{file=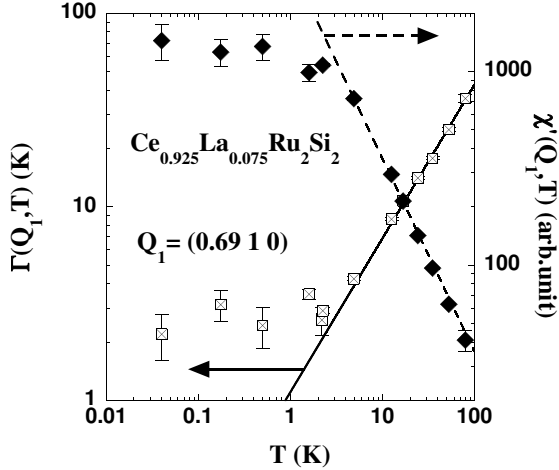,width=73mm}
    \caption{Temperature dependence of the relaxation rate $\Gamma(\mathbf{Q}_{1},T)$
     and the static susceptibility $\chi'(\mathbf{Q}_{1},T)$ of
     Ce$_{0.925}$La$_{0.075}$Ru$_{2}$Si$_{2}$ \cite{Knafo1,Knafo2}.}
    \label{largint}
\end{figure}

\begin{figure}[t!]
    \centering
    \epsfig{file=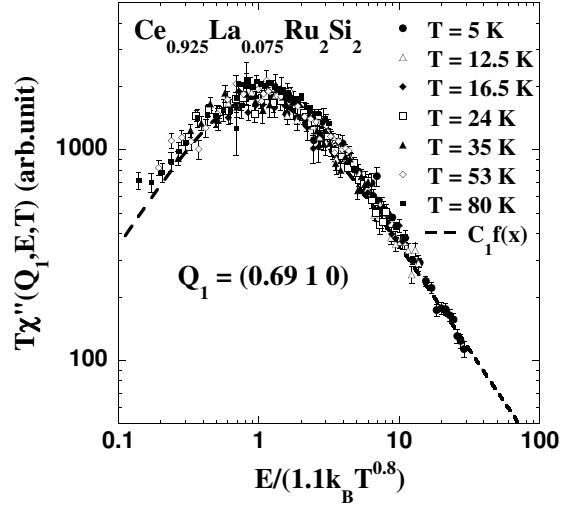,width=73mm}
    \caption{Scaling behavior of the antiferromagnetic
    fluctuations of Ce$_{0.925}$La$_{0.075}$Ru$_{2}$Si$_{2}$,
    obtained for $5\leq T\leq80$ K at the momentum transfer
    $\mathbf{Q}_{1}$ \cite{Knafo1,Knafo2}.} \label{scaling}
\end{figure}

We recall here results concerning the antiferromagnetic
fluctuations of the heavy-fermion compound
Ce$_{0.925}$La$_{0.075}$Ru$_{2}$Si$_{2}$
\cite{Knafo1,Knafo2,Knafo3}. These excitations are measured at a
momentum transfer $\mathbf{Q}_{1}=(0.69,1,0)$, which corresponds
to a correlated signal with the wavevector $\mathbf{k}_{1}$ =
(0.31,0,0). Knowing that the scattered intensity is proportional
to the scattering function $S(\mathbf{Q},E,T)$, the imaginary part
of the dynamical susceptibility $\chi''(\mathbf{Q},E,T)$ is
deduced using the fluctuation-dissipation theorem:
\begin{eqnarray}
  S(\mathbf{Q},E,T) &=&
  \frac{1}{\pi}\frac{1}{1-e^{-E/k_{B}T}}\chi''(\mathbf{Q},E,T).
\end{eqnarray}
Then, the susceptibility is fitted with a single quasielastic
Lorentzian shape:
\begin{eqnarray}
  \chi''(\mathbf{Q}_{1},E,T) &=&
  \chi'(\mathbf{Q}_{1},T)\frac{E/\Gamma(\mathbf{Q}_{1},T)}{1+(E/\Gamma(\mathbf{Q}_{1},T))^{2}},
  \label{lor}
\end{eqnarray}
where $\chi'(\mathbf{Q}_{1},T)$ and $\Gamma(\mathbf{Q}_{1},T)$ are
respectively the static susceptibility and the relaxation rate of
the antiferromagnetic fluctuations. The variations with $T$ of
those two parameters are plotted in Figure \ref{largint}. We
obtain that the dynamical susceptibility depends weakly on $T$ for
$T<T_1$, with $T_1=2.5$ K. In this regime, the static
susceptibility and the relaxation rate are given approximately by:
\begin{eqnarray}
\chi'(\mathbf{Q}_{1},T)=\frac{C_1}{T_{1}}\;\;\;\;\;\;{\rm{and}}\;\;\;\;\;\;\Gamma(\mathbf{Q}_{1},T)=
k_{B}T_{1}.
\end{eqnarray}
For $T>T_1$, the susceptibility becomes controlled by $T$ and we
have:
\begin{eqnarray}
\chi'(\mathbf{Q}_{1},T)=\frac{C_1}{T}\;\;\;\;{\rm{and}}\;\;\;\;\Gamma(\mathbf{Q}_{1},T)=
a_{1}T^{0.8}. \label{ht}
\end{eqnarray}
This is only in this high temperature regime that
antiferromagnetic fluctuations follow a scaling law: using
(\ref{lor}) and (\ref{ht}), we obtain immediately:
\begin{eqnarray}
T\chi''(\mathbf{Q}_{1},E,T) = C_{1}f[E/(a_{1}T^{0.8})],
\end{eqnarray}
where $f(x) =x/(1+x^{2})$ and $x=E/(a_{1}T^{0.8})$. In Figure
\ref{scaling}, this scaling behavior is plotted for temperatures
$T>T_1$ and all the data collapse effectively on a single curve.
This law does not enter the framework of usual quantum phase
transition theories, where $\beta$ cannot be smaller than 1
\cite{Millis,Moriya,Continentino}. The reason is that those
theories do not consider the temperature dependence of the Kondo
local magnetic fluctuations. In the studies reported in References
\cite{Knafo1,Knafo2}, we also showed that it is necessary to
consider the spectra at different wavevectors corresponding
respectively to antiferromagnetic fluctuations and to local
magnetic fluctuations.

\section{Artificial scaling laws}

Several studies of heavy-fermion systems at their quantum phase
transition report different kinds of scaling laws of the general
form
\cite{Aronson95,Schroder00,Park02a,Park02b,Montfrooij03,So03,Knafo1,Knafo2}:
\begin{eqnarray}
  T^{\alpha}\chi''(E,T)=f(E/T^\beta).
  \label{scalgen}
\end{eqnarray}
In our study of Ce$_{0.925}$La$_{0.075}$Ru$_{2}$Si$_{2}$, each
spectrum is analyzed separately at a given momentum transfer
$\mathbf{Q}$ and at a given temperature $T$ . Then, for the
antiferromagnetic fluctuations measured at the wavevector
$\mathbf{Q}_1$, we deduced the exponents $\alpha=1$ and
$\beta=0.8$ from the temperature dependance of the static
susceptibility and of the relaxation rate. In the literature, the
scaling laws are often obtained as follows: $\beta$ is fixed to 1
and $\alpha$ is chosen for the best collapse of the data on a
single curve when $T^{\alpha}\chi''(E,T)$ is plotted against
$E/T^\beta$. We show here how such a method can lead to erroneous
results and to contradictions with the real physics of the system.
In fact, for each value of $\beta$ it is possible to obtain an
optimal value of $\alpha$ so that the data collapse in a single
curve, this artificial scaling law being sometimes restricted to
some ranges
of temperatures and/or energies.\\

\begin{figure}[b!]
    \centering
    \epsfig{file=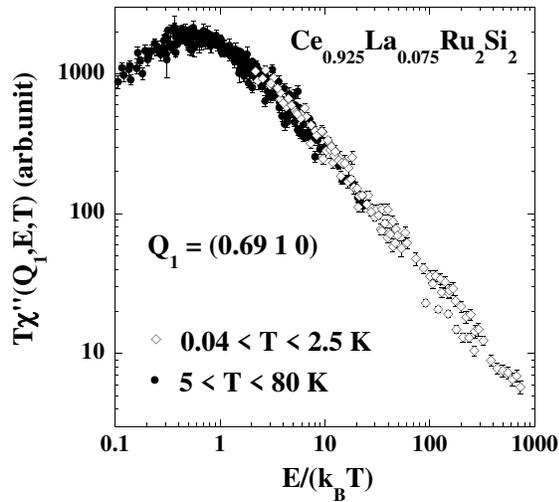,width=73mm}
    \caption{Variation of $T^{\alpha}\chi''(\mathbf{Q}_{1}E,T)$ in
    function of
    $E/T^{\beta}$ for Ce$_{0.925}$La$_{0.075}$Ru$_{2}$Si$_{2}$:
    $\beta$ is fixed to 1 and $\alpha=1$ corresponds to the best
    collapse of the data on a single curve.} \label{scaling2}
\end{figure}

\begin{figure}[t!]
    \centering
    \epsfig{file=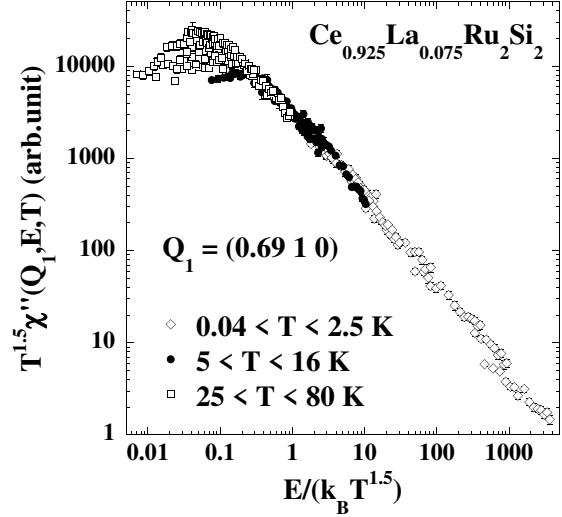,width=73mm}
    \caption{Variation of $T^{\alpha}\chi''(\mathbf{Q}_{1}E,T)$ in
    function of
    $E/T^{\beta}$ for Ce$_{0.925}$La$_{0.075}$Ru$_{2}$Si$_{2}$:
    $\beta$ is fixed to 1.5 and $\alpha=1.5$ corresponds to the best
     collapse of the data on a single curve.} \label{scaling3}
\end{figure}

Let us consider the example of the antiferromagnetic fluctuations
of Ce$_{0.925}$La$_{0.075}$Ru$_{2}$Si$_{2}$. Figures
\ref{scaling2} and \ref{scaling3} correspond to the plot of
$T^{\alpha}\chi''(\mathbf{Q}_{1},E,T)$ against $E/T^{\beta}$,
where $\beta$ is fixed and $\alpha$ is chosen for the best
collapse of the data on a single curve. We see in Figure
\ref{scaling2} that when $\beta=1$, the best collapse of the data
is obtained for $\alpha=1$ and is followed for temperatures from
80 K down to 40 mK. In Figure \ref{scaling3}, we fix $\beta=1.5$
and the best collapse of the data is found for $\alpha=1.5$,
corresponding to temperatures from 16 K to 40 mK. In this second
graph, a collapse of the data on a single curve is not obtained for $T\geq25$ K.\\

From the results summarized in Section 2, we know that the
antiferromagnetic dynamical susceptibility is almost $T$-
independent for $T<T_1$, and consequently that this low
temperature regime can not correspond to any scaling law in
$E/T^\beta$. The plots of Figures \ref{scaling2} and
\ref{scaling3} correspond thus to artificial scaling laws, the
collapse of the low temperature data in a single curve coming not
from some intrinsic physics. In fact, the regime $T<T_{1}$ is
characterized by the nearly $T$-independent dynamical
susceptibility:
\begin{eqnarray}
  \chi''(\mathbf{Q}_{1},E,T) &=&
  \chi'(\mathbf{Q}_{1},0)\frac{E/\Gamma(\mathbf{Q}_{1},0)}{1+(E/\Gamma(\mathbf{Q}_{1},0))^{2}}.
  \label{lorquasielfit2}
\end{eqnarray}
Knowing that $\Gamma(\mathbf{Q}_{1},0)\simeq k_{B}T_{1}\simeq0.3$
meV and that our data correspond to energy transfers $0.4<E<9.5$
meV, we can make the approximation $E\gg\Gamma(\mathbf{Q}_{1},0)$,
so that:
\begin{eqnarray}
  \chi''(\mathbf{Q}_{1},E,T) &=&
  \chi'(\mathbf{Q}_{1},0)\left(\frac{E}{\Gamma(\mathbf{Q}_{1},0)}\right)^{-1}\sim\left(E\right)^{-1}.
  \label{lorquasielfit3}
\end{eqnarray}
The dynamical susceptibility can then be written as:
\begin{eqnarray}
  T^{\alpha}\chi''(\mathbf{Q}_{1},E,T)
  \sim\left(\frac{E}{T^{\alpha}}\right)^{-1},
  \label{lorquasielfit4}
\end{eqnarray}
which means that an optimal collapse of the data measured at the
temperatures $T<T_{1}$ will be obtained for $\alpha=\beta$, when
$T^{\alpha}\chi''(\mathbf{Q}_{1},E,T)$ is plotted against
$E/T^{\beta}$. In a $log$-$log$ plot of
$\chi''(\mathbf{Q}_{1},E,T)$ as a function of $E$, the almost
$T$-independent spectrum which is obtained for $T<T_1$ corresponds
to a segment of a straight line of slope -1. When
$T^{\alpha}\chi''(\mathbf{Q}_{1},E,T)$ is plotted against
$E/T^{\alpha}$, it leads to a shift of this segment on the same
straight line. This simple geometric construction
is the only origin of the artificial scaling.\\

Figures \ref{scaling2} and \ref{scaling3} correspond effectively
to two cases of artificial scaling laws of the form
(\ref{lorquasielfit4}). Moreover, those erroneous laws are not
only obtained for the low temperature data, but also for parts of
the high temperature data. Indeed, Figure \ref{scaling2} is
characterized for $T>T_1$ by a collapse of the data in a single
curve, since the values $\alpha=1$ and $\beta=1$ are very close to
the correct values $\alpha=1$ and $\beta=0.8$. In Figure
\ref{scaling3}, the exponents $\alpha=1.5$ and $\beta=1.5$ are
quite different from the correct values $\alpha=1$ and
$\beta=0.8$. As a consequence, the artificial scaling law is
obtained only up to 16 K and for $T\geq25$ K, the data do not
collapse any more in a single curve. We recall that the only
$E/T^{\beta}$ scaling behavior of the antiferromagnetic
fluctuations of Ce$_{0.925}$La$_{0.075}$Ru$_{2}$Si$_{2}$
corresponds to the temperature regime $T>T_1$, with the exponents
$\alpha=1$ and $\beta=0.8$, as explained in Section 2.\\

Other kinds of artificial scaling laws can also be obtained for
non Lorentzian spectra: at sufficiently small temperatures, when
the dynamical magnetic susceptibility is nearly $T$-independent
such as, instead of (\ref{lorquasielfit3}):
\begin{eqnarray}
  \chi''(\mathbf{Q}_{1},E,T) \sim\left(E\right)^{-\gamma},
  \label{lorquasielfit5}
\end{eqnarray}
we obtain immediately:
\begin{eqnarray}
  T^{\alpha}\chi''(\mathbf{Q}_{1},E,T)
  \sim\left(\frac{E}{T^{\alpha/\gamma}}\right)^{-\gamma}.
  \label{lorquasielfit6}
\end{eqnarray}
An artificial scaling law can thus be plotted with
$\beta=\alpha/\gamma$. When $\beta$ is fixed, a collapse of the
low temperature data in a single curve is consequently obtained
for an exponent $\alpha=\beta\gamma$. This signifies that, when
$\beta$ is fixed to 1, an artificial scaling law of the form
(\ref{scal}) is obtained with the exponent $\alpha=\gamma$.\\

Finally, depending on the temperature and energy ranges which are
considered, but also on the shape of the spectra, several kinds of
erroneous scaling laws can be plotted. Equations
(\ref{lorquasielfit4}) and (\ref{lorquasielfit6}) are just two
examples of such artificial scaling laws. We illustrate the former
case using the graphs of the Figures \ref{scaling2} and
\ref{scaling3}, where $\alpha=1$ and $\alpha=3/2$, respectively.
Those two graphs could be taken as proofs of diverging
fluctuations for $T \rightarrow0$. Moreover, the plot of Figure
\ref{scaling3} could be abusively interpreted as a verification of
quantum phase transition theories for an itinerant system
\cite{Millis,Moriya,Continentino}. Actually, for the spatial
dimension $d=3$ and the critical exponent $z=2$ corresponding to
Ce$_{1-x}$La$_{x}$Ru$_{2}$Si$_{2}$, such models predict i) that
the anomalous exponents of the scaling law (\ref{scalgen}) are
$\alpha=3/2$ and $\beta=3/2$, ii) that the fluctuations should
diverge for $T\rightarrow0$ and iii) that this law would not been
verified for temperatures bigger than the Kondo temperature, which
is here equal to 18 K \cite{Knafo1,Knafo2}. However, we know that
this is not verified in Ce$_{0.925}$La$_{0.075}$Ru$_{2}$Si$_{2}$
since the fluctuations saturate below $T_{1}=2.5$ K and thus no
scaling can be obtained in the low temperature regime of the
magnetic fluctuations. This illustrates perfectly the dangers of a
unique graphical determination of scaling laws of the form
$T^\alpha\chi"(E,T)=g(E/T^\beta)$.

\section{Conclusion}

In this letter, we used the specific example of the
antiferromagnetic fluctuations of
Ce$_{0.925}$La$_{0.075}$Ru$_{2}$Si$_{2}$ to show how erroneous
$E/T^\beta$ scaling laws of the dynamical magnetic susceptibility
can be plotted. Such artificial scaling laws are obtained when the
exponents $\alpha$ and $\beta$ are only determined graphically, as
done in Section 3, and come just from a simple geometric
construction. This can lead to misunderstand the physics of the
system; the erroneous scaling laws shown here are plotted without
any divergence of the fluctuations, while such $E/T^\beta$ scaling
laws obtained down to $T=0$ should be associated to diverging
fluctuations. We stress thus that a scaling behavior can only be
established after a precise study of the temperature dependence of
the dynamical magnetic susceptibility and of the parameters which
characterize it, as shown in Section 2. A great care has also to
be given to the $\bf{Q}$-dependence of those spectra and to the
corresponding scaling laws \cite{Knafo1,Knafo2}.

\end{document}